\documentclass[sigconf,natbib=false]{acmart}

\setcctype{by}
\renewcommand\footnotetextcopyrightpermission[1]{}

\usepackage{tabularx}
\usepackage{array}
\usepackage{multirow}
\usepackage{booktabs}
\usepackage{makecell}
\usepackage{adjustbox}
\usepackage{svg}
\usepackage{makecell}
\usepackage{enumitem}

\RequirePackage[
  datamodel=acmdatamodel,
  style=acmnumeric,
  ]{biblatex}

\begin{document}
\pagestyle{plain}

\title{An Autonomy-Aware Metamodel for Human–AI Collaboration in Software Engineering}

\author{Md Mahade Hasan}
\email{mdmahade.hasan@tuni.fi}
\orcid{0009-0006-5154-7331}
\affiliation{%
  \institution{Tampere University}
  \country{Finland}
}

\author{Md Toufique Hasan}
\email{mdtoufique.hasan@tuni.fi}
\orcid{0009-0005-8907-9898}
\affiliation{%
  \institution{Tampere University}
  \country{Finland}
}

\author{Zheying Zhang}
\email{zheying.zhang@tuni.fi}
\orcid{0000-0002-6205-4210}
\affiliation{%
  \institution{Tampere University}
  \country{Finland}
}

\author{Kalle Kulonen}
\email{kalle.kulonen@tuni.fi}
\orcid{0009-0007-6734-6734}
\affiliation{%
  \institution{Tampere University}
  \country{Finland}
}

\author{Konsta Kalliokoski}
\email{konsta.m.kalliokoski@jyu.fi}
\affiliation{%
  \institution{University of Jyväskylä}
  \country{Finland}
}

\author{Fran\c{c}ois Christophe}
\email{francois.m.christophe@jyu.fi}
\affiliation{%
  \institution{University of Jyväskylä}
  \country{Finland}
}

\author{Tommi Mikkonen}
\email{tommi.j.mikkonen@jyu.fi}
\affiliation{%
  \institution{University of Jyväskylä}
  \country{Finland}
}

\author{Pekka Abrahamsson}
\email{pekka.abrahamsson@tuni.fi}
\orcid{0000-0002-4360-2226}
\affiliation{%
 \institution{Tampere University}
 \country{Finland}
}


\renewcommand{\shortauthors}{Hasan et al.}

\begin{abstract}
Artificial Intelligence (AI) is shifting software engineering from tool-supported processes towards AI-first collaboration, 
where authority is dynamically distributed across human and artificial actors. However, 
existing method engineering approaches
assume static, human-centric control and provide limited support 
explicitly capturing evolving autonomy. This paper presents a vision for autonomy-aware method engineering by proposing a metamodel that 
treats autonomy not as a fixed property of an actor, but as a derived, situation-dependent authority assignment determined by task, context, and collaboration pattern.
The metamodel formalizes autonomy through four authority 
dimensions: task execution, task decomposition, task initiation, and collaboration reconfiguration. 
Through an analytical instantiation with a multi-agent requirements analysis tool, we illustrate how the metamodel supports dynamic authority assignment. 
This work provides a conceptual foundation for governance-aware, adaptable, and AI-first software engineering methods.

\end{abstract}




\maketitle

\section{Introduction}

Software engineering (SE) has evolved through successive waves of abstraction, transitioning from informal coordination and manual artifact production towards increasingly formalized and tool-supported practices. From 
Computer-Aided Software Engineering (CASE) environments to method engineering approaches based on process metamodels, 
such as the Software and Systems Process Engineering Metamodel (SPEM), these advances established modeling backbones that made roles, activities, and artifacts governable, and enabled systematic construction and tailoring of SE methods to project contexts \cite{fuggetta2002classification, sommerville2016software, OMG2008, brinkkemper1996method, henderson2010situational}.


The growing integration of AI in SE marks a new transition. Development practices are 
shifting toward increasingly human–AI collaborative teams, where AI agents not only generate artifacts but also execute tasks, decompose work, and coordinate activities. In contrast to earlier tool-assisted environments with 
stable human-centered authority, contemporary AI systems increasingly participate in task allocation, workflow configuration, and decision-making \cite{hassan2024towards, naik2025exploring}. This shift introduces 
dynamic and distributed authority, where control varies across human and artificial actors, tasks, and contexts, and evolves with advancing AI capabilities.

However, existing 
method engineering approaches largely assume static authority assignments and provide limited support for representing dynamic autonomy and situational authority redistribution. While they effectively capture process structure, 
they do not explicitly model how autonomy varies across collaboration settings or adapts to context and governance constraints. 
Consequently, the transition from tool-assisted workflows toward AI-first SE teams remains conceptually under-specified and methodologically unguided. Recent work on multi-agent tools \cite{ait2026, ait2025towards, MGS-2012} explores autonomous workflows but typically treats autonomy as implicit within execution models rather than as an explicit modeling dimension.

To address this gap in method engineering, this study proposes an autonomy-aware metamodel for modeling dynamic collaboration in AI-first SE teams. The metamodel defines seven core constructs -- Actor, Task, Capability, Artifact, Collaboration Pattern, Constraint, and Context -- and treats autonomy as a derived, situation-dependent property of Actors instances. Specifically, an actor’s autonomy level is determined by the interplay of task, context, and collaboration pattern. This enables flexible authority assignment while preserving a stable process structure, and makes coordination and governance explicit through collaboration patterns and interaction rules.

This work is visionary in its nature. Through analytical instantiation with multi-agent requirements analysis tool \cite{req_tool}, we illustrate how the metamodel can support variation in autonomy assignments and authority redistribution. 
By making autonomy and collaboration explicit at the metamodel level, the paper contributes a foundation for future research on adaptable SE methods in AI-first teams.



The remainder of this paper is organized as follows. Section~\ref{sec:background} reviews related work. Section~\ref{sec:metamodel} presents the proposed metamodel. Section~\ref{sec:instantiation} illustrates its use through an analytical instantiation. Section~\ref{sec:discussion} discusses implications and future work, and Section~\ref{sec:conclusion} concludes the paper.

\section{Background and Related Work}
\label{sec:background}
Method engineering provides the foundation for constructing, adapting, and analyzing SE methods. Within method engineering, process metamodeling provides a formal foundation for representing methods as explicit, analyzable, and reusable structures.
Established frameworks such as SPEM 2.0~\cite{OMG2008}, ISO/IEC 24744~\cite{gonzalez2007supporting}, and the SEMAT Essence kernel~\cite{jacobson2012essence} define abstractions for roles, tasks, activities, work products, and their relationships. They enable reuse, configuration, consistency checking, and reasoning over method structures~\cite{brinkkemper1996method,inproceedingsHarmsen1994,henderson2010situational}. Yet these method engineering approaches largely assume static, human-centered method structures and provide limited support for modeling how authority, coordination, and decision-making are distributed across human and AI actors.

\noindent\textbf{Ontology and Agent-Oriented Approaches.}
Ontology-based and agent-oriented approaches complement process metamodeling by formally representing systems, artifacts, and interactions. Unified design and agent-oriented ontologies capture goals, tasks, behaviors, and commitments for semantic interoperability and automated reasoning across heterogeneous systems~\cite{lu2021design,bella2023ontology,bauer2004methodologies}. Process-oriented agent engineering has also used BPMN to model high-level agent behavior and communication in multi-agent systems~\cite{MGS-2012}. LLM-based ontology engineering uses agent-based architectures combining structured metadata (symbolic) and text embeddings (vector) of ontology entities to improve semantic matching and reasoning~\cite{qiang2023agent}. Foundation-model-based agents have also been studied through design patterns for planning, reflection, multi-agent cooperation, and guardrail mechanisms~\cite{liu2025agent}. However, these approaches mainly model predefined system behavior and interaction patterns, rather than the construction and adaptation of SE methods. Consequently, they enrich semantic rigor and provide useful architectural and process-oriented guidance, but do not provide a process metamodel in which methodological structure and authority distribution are modeled together.

\noindent\textbf{Human-AI Collaboration in Software Engineering.}
Advances in AI are transforming SE into human-AI collaboration, where AI systems participate in task execution, decomposition, and coordination. Emerging multi-agent development frameworks~\cite{hassan2024towards,hong2023metagpt,ge2025survey} conceptualize development as coordinated systems of human and artificial agents, where specialized AI agents assume roles, exchange artifacts, and iteratively refine outputs. Recent multi-agent SE systems further illustrate this shift by distributing responsibilities among specialized AI agents that collaboratively analyze code, perform refactoring tasks, and iteratively improve software artifacts~\cite{10.1007/978-3-032-12089-2_26}. Empirical studies show that developers increasingly perceive these systems as collaborative teams operating across a spectrum from AI-assisted to more autonomous workflows~\cite{naik2025exploring}.

This transformation introduces challenges for coordination, transparency, and governance. Interacting agents can cause error propagation, non-deterministic behavior, and limited interpretability, motivating layered transparency and structured communication for debugging and control~\cite{naik2025exploring,ge2025survey}. Recent work on automated governance for human-agent collaboration in software projects argues that joint decision-making by humans and AI-powered agents requires explicit mechanisms for assigning roles and responsibilities, as well as for establishing consensus among heterogeneous participants~\cite{ait2025towards}. Complementing this perspective, agent design patterns further highlight the importance of making planning, cooperation, team structure, and decision mechanisms explicit in agentic systems~\cite{liu2025agent}. Hybrid human-AI teams also redistribute effort across the lifecycle, where the actual and perceived autonomy influence communication, coordination, and shared mental models, while task decomposition and role allocation improve collaboration and situational awareness~\cite{musick2021happens,karakama2021task}.

\noindent\textbf{Toward Autonomy-Aware Method Models.}
Together, these research streams motivate autonomy-aware method modeling, but they remain fragmented. Method engineering structures development methods, ontology and agent-oriented work models system semantics and interaction, and AI-driven SE studies collaboration. None provides a unified process metamodel for authority redistribution under varying autonomy conditions. From a governance perspective, Domain Specific Language (DSL) for human-agent collaboration treats autonomy as an explicit AI-agent characteristic, but does not define autonomy levels or explain how different degrees of autonomy redistribute authority across development~\cite{ait2025towards}. Prior work has examined autonomy levels in human-autonomy teaming~\cite{HumanAutonomy}, and AI-SEAL~\cite{Feldt} uses automation level to classify AI applications in SE and reason about risk. Recent BPMN work also supports human-agentic workflow representation~\cite{ait2026}. Overall, these works view autonomy mainly through interaction, workflow, governance, or risk-analysis lenses, rather than as a method-modeling construct for explaining authority redistribution across actors, tasks, constraints, contexts, and collaboration patterns. As a result, there remains a lack of a SE process metamodel that captures both method engineering structure and dynamic autonomy redistribution in human-AI collaboration.

This gap motivates an autonomy-aware metamodel that models autonomy as a first-class dimension for human-AI SE, allowing authority distribution to vary across human and artificial actors without redesigning the underlying process structure. Unlike workflow-level views, this approach treats autonomy as a method-engineering concern for specifying and adapting responsibility and authority.

\begin{figure*}[t]
    \centering
    \includegraphics[width=\textwidth,keepaspectratio]{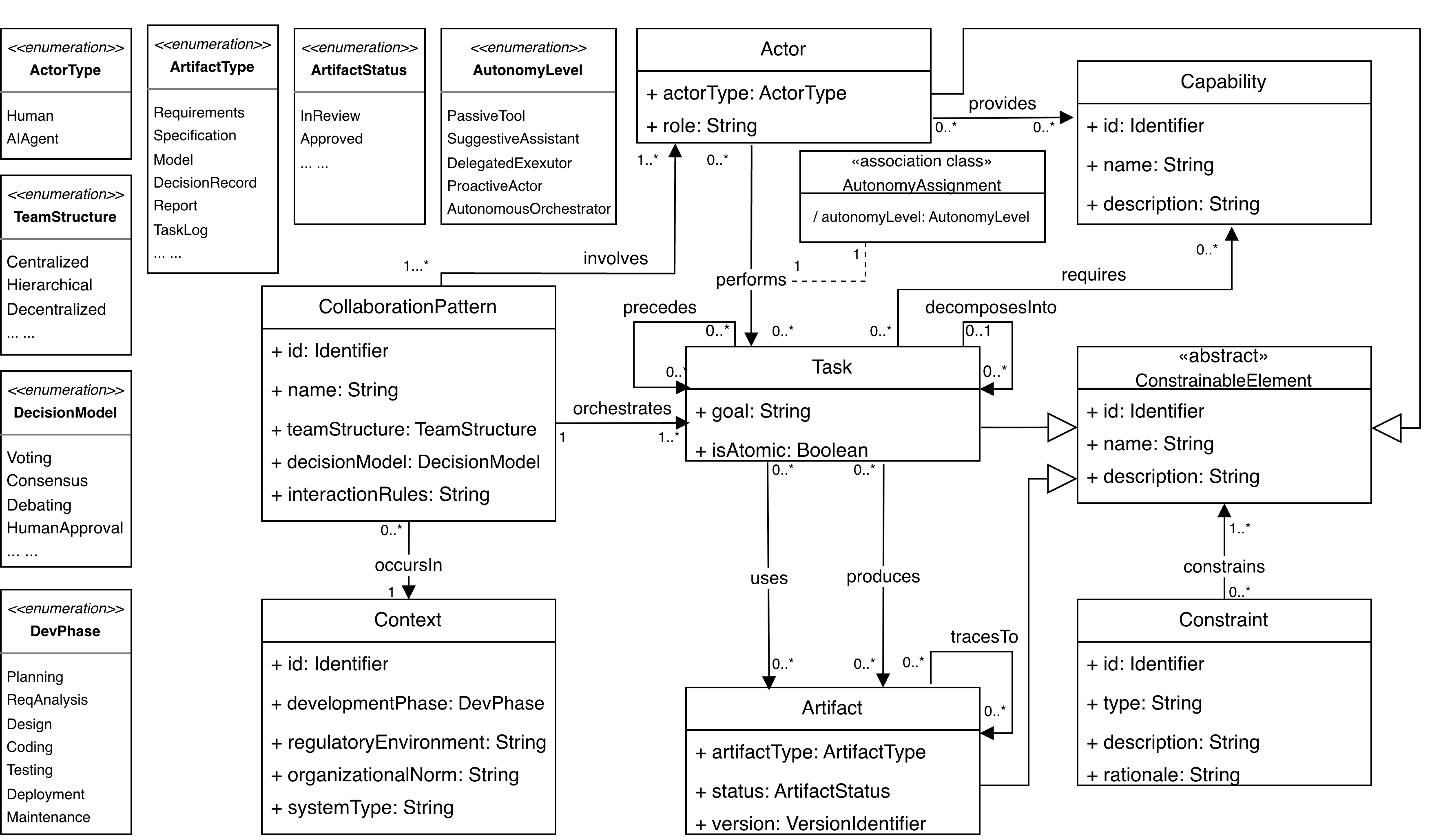}
    \caption{A metamodel specifying human-AI collaborative software engineering methods}
    \label{fig:meta_model}
\end{figure*}

\section{Autonomy-Aware Metamodel Definition}
\label{sec:metamodel}
Building on the limitations identified in the previous section, this study proposes an autonomy-aware metamodel for human-AI collaborative SE. In this work, autonomy-aware denotes that autonomy is modeled as an explicit attribute of actors, capturing the degree of authority they hold over task execution, task decomposition, task initiation, and collaboration reconfiguration. By representing autonomy at the actor level, the metamodel separates the structural definition of a process from the distribution of authority among human and AI actors. As a result, autonomy can vary without changing the core structure of the method. Grounded in method engineering approaches, the metamodel represents human and artificial participants uniformly as actors embedded in a shared socio-technical method structure. This allows the metamodel to represent both the stable structure of a collaborative method and the redistribution of authority as human-AI practices evolve toward higher-autonomy.
Figure~\ref{fig:meta_model} presents the metamodel as a class diagram specifying its core constructs, attributes, and relationships.

\subsection{Core Constructs}

The metamodel comprises seven core constructs: \textit{Actor}, \textit{Task}, \textit{Capability}, \textit{Artifact}, \textit{CollaborationPattern}, \textit{Constraint}, and \textit{Context}. These constructs capture who participates in human-AI collaborative SE practices, what activities are performed, which capabilities and artifacts are involved, under which contextual conditions collaboration occurs, and which constraints govern the process. In addition, the abstract class \textit{ConstrainableElement} serves as a common superclass for \textit{Actor}, \textit{Task}, and \textit{Artifact}. This allows constraints to be uniformly attached to these elements. This section defines each construct and its role in the metamodel.

An \textbf{Actor} is a participant in performing a SE task. It can be either a human or an AI agent, as specified by the \textit{actorType} attribute. Actors are represented uniformly at the metamodel level, while their operational authority is distinguished through the \textit{autonomyLevel}, a dynamic and context-dependent attribute. This attribute defines the extent to which an actor may execute, decompose, or initiate tasks, or modify collaboration structures. Thus, authority becomes an explicit and configurable modeling parameter rather than an implicit property of the process. Each actor also has a \textit{role} describing its responsibility in the method and may provide one or more \textit{Capability} instances.

A \textbf{Task (T)} represents a distinct piece of SE work within the method. It may be atomic or decomposed into sub-tasks, thereby supporting the hierarchical structuring of activities. Tasks may also be ordered through the \textit{precedes} relation, which indicates that one task should occur before another. \textit{Task}s are performed by \textit{Actor}s with the required \textit{Capability} instances and are linked to the \textit{Artifact}s they use or produce. They are orchestrated through \textit{CollaborationPattern}s, which specify how work is coordinated across actors. Thus, \textit{Task} provides an explicit unit for modeling execution, decomposition, coordination, and responsibility allocation. The \textit{goal} attribute describes the intended outcome of the task.

A \textbf{Capability (Cap\textit{)}} represents a functional ability needed to perform a task. In the metamodel, \textit{Task}s are associated with one or more \textit{Capability} instances they require, while \textit{Actor}s possess \textit{Capability} instances that enable task performance. This separates task requirements from performer identity, and allows task allocation to be modeled in terms of required abilities rather than fixed roles. As a result, the metamodel supports substitution and gradual redistribution of responsibilities between human and AI \textit{Actor}s.

An \textbf{Artifact (A)} is a work product created, used, modified, or maintained during the execution of a \textit{Task}. Artifacts include outputs such as requirements documents, design models, source code, test cases, and deployment scripts, as well as intermediate items relevant to the software system or its development. They are versioned and may trace to other \textit{Artifact} instances. Artifacts provide the basis for tracing, coordination, and evaluation across \textit{Task}s and \textit{Actor}s. 

A \textbf{CollaborationPattern} encapsulates a reusable coordination structure that organizes actors, tasks, and artifacts into a coherent method fragment. It specifies how task is coordinated through \textit{interactionRules}, how participating actors are organized through \textit{teamStructure} (e.g., centralized, hierarchical, or decentralized), and how decisions are reached through \textit{decisionModel} (e.g., Voting,  Debating, Consensus, or HumanApproval ). By making coordination, team structures, and decision-making explicit, a CollaborationPattern provides a reusable specification of both collaboration and governance in human–AI SE  methods. Each CollaborationPattern is situated within a particular Context, which defines its situational applicability and normative boundaries.

A \textbf{Constraint (C)} specifies rules or conditions that restrict or regulate the permissible behavior of elements in a method. Constraints are prescriptive elements of the metamodel: they define what is allowed, required, or prohibited for constrainable elements. They may capture governance policies, validation requirements, compliance obligations, safety conditions, or organizational procedures. For example, a governance rule may require human approval before a deployment task, a safety condition may restrict an AI actor from autonomously executing critical tasks, and a validation rule may require that a generated artifact satisfy predefined quality criteria. Constraints are connected to \textit{ConstrainableElement}s through the \textit{constrains} relation, allowing the same constraint mechanism to apply to \textit{Actor}s, \textit{Task}s, and \textit{Artifact}s.

A \textbf{Context (Cx)} characterizes the situational environment in which a \textit{CollaborationPattern} is applicable. Unlike Constraints, which prescribe rules or restrictions, Context provides descriptive information about the operating environment. It has the attributes \textit{id}, \textit{developmentPhase}, \textit{regulatoryEnvironment}, \textit{organizationalNorm}, and \textit{systemType}, which capture lifecycle, regulatory, organizational, and system-specific conditions. For example, a task in a highly regulated domain may require a different collaboration pattern than the same task in an exploratory prototype setting. By separating situational applicability from collaboration structure, the metamodel supports reuse and adaptation across different environments.

Together, the above constructs provide a stable basis for modeling authority redistribution.

\begin{table*}[htbp]
\centering
\caption{Autonomy Levels with Operational Authority and Alignment to Related Classifications}
\label{tab:autonomy_levels}
\small
\setlength{\tabcolsep}{4pt}
\resizebox{\textwidth}{!}{
\begin{tabular}{@{}l l l l l l l l@{}}
\toprule
Level & Description
& Task Execution
& Task Decomposition
& Task Initiation
& Modify Collaboration
& Human Decision Role
& LA scale~\cite{Sheridan,Parasuraman,Feldt} \\
\midrule

AL0 & PassiveTool
& 0
& 0
& 0
& 0
& Always
& LA1 \\

AL1 & SuggestiveAssistant
& 0 (suggestion)
& 0
& 0
& 0
& Required
& LA2--LA4 \\

AL2 & DelegatedExecutor
& 1
& 0
& 0
& 0
& Context-dependent
& LA5--LA6 \\

AL3 & ProactiveActor
& 1
& 1
& 1
& 0
& Required for critical decision
& LA6--LA7 \\

AL4 & AutonomousOrchestrator
& 1
& 1
& 1
& 1
& Human oversight
& LA8--LA10 \\

\bottomrule
\end{tabular}
}
\end{table*}

\subsection{Autonomy Levels}
\label{sec:AutonomyLevels}

Autonomy constitutes the central differentiating dimension of the proposed metamodel. Rather than embedding authority assumptions within Actor types, autonomy is modeled explicitly as a parameter that governs an actor's authority over tasks and collaboration patterns. This makes authority redistribution analyzable and allows human-AI collaboration to vary without changing the structure of the method. Formally, for a given actor in a specific situation, autonomy can be represented as:
\[
\mathit{AutonomyLevel} = f(\mathit{Task}, \mathit{Context}, \mathit{CollaborationPattern})
\]
This formulation emphasizes that autonomy is not an inherent or fixed property of an actor. The same actor may receive different autonomy level assignments depending on the authority required by the task, the situational restrictions imposed by the context, and the coordination and decision rules defined by the collaboration. 

\subsubsection{Rationale for Autonomy Specification}

Autonomy is defined in terms of operational authority over four dimensions of process control. These dimensions correspond to the minimal set of authority boundaries that distinguish meaningfully different forms of control in SE  practices: executing an assigned task, structuring work independently, initiating further action, and modifying the collaboration itself. These dimensions provide the basis for defining the autonomy levels from AL0 to AL4. Together, they form a complete and non-redundant characterization of operational independence in human-AI collaboration:

\begin{enumerate}
\item \textbf{Task Execution} -- authority to perform a defined task.
\item \textbf{Task Decomposition} -- authority to refine a task into sub-tasks.
\item \textbf{Task Initiation} -- authority to initiate new tasks without an external trigger.
\item \textbf{Collaboration Reconfiguration} -- authority to modify participating actors, workflow structure, artifact flow, contextual bindings, or applicable constraints.
\end{enumerate}

Across these four dimensions, autonomy captures the extent to which an actor can autonomously sense, decide, and act to achieve task-specific goals~\cite{RobotAutonomy}. This aligns with prior work that characterizes autonomy as graded decision latitude over information acquisition, decision selection, and action implementation in human-automation and human-autonomy teaming~\cite{HumanAutonomy}. It also corresponds to established classifications of automation levels in human-machine collaboration~\cite{Sheridan,Parasuraman,Feldt}.

\subsubsection{Autonomy Level Definition}

Grounded in the four authority dimensions defined above, we define five autonomy levels (ALs), as summarized in Table~\ref{tab:autonomy_levels}. These levels characterize progressively broader distributions of control between human and AI actors.

\begin{itemize}
    \item AL0: PassiveTool. The actor has no authority over task execution, task decomposition, task initiation, or collaboration reconfiguration; all such authority remains with human actors.
    \item AL1: SuggestiveAssistant. The actor may generate suggestions or other supportive materials, but it does not execute tasks or produce the primary output artifacts of the task. Human or AI actors remain responsible for producing the output artifacts. Any operational use of AL1 outputs requires human confirmation.
    \item AL2: DelegatedExecutor. The actor may execute assigned tasks within predefined boundaries. However, authority over task decomposition, task initiation, and collaboration structure remains with human actors.
    \item AL3: ProactiveActor. The actor may execute tasks, decompose them into sub-tasks, and initiate sub-tasks within the assigned scope of work. Critical decisions remain subject to human approval.
    \item AL4: AutonomousOrchestrator. The actor may execute tasks, decompose work, initiate new tasks, and modify collaboration patterns. Human involvement shifts from direct control to supervisory oversight.
 \end{itemize} 

Taken together, these levels provide a structured basis for modeling gradual authority redistribution from human-centered methods to AI-first orchestration. To situate them with respect to prior work, Table~\ref{tab:autonomy_levels} relates each autonomy level to indicative ranges in the Levels of Automation (LA) scale, originally proposed by Sheridan and Verplank~\cite{Sheridan}. The LA scale ranges from LA1, where the human makes and implements all decisions, to LA10, where the computer acts fully autonomously and informs the human only if it decides to do so. The LA scale was extended by Parasuraman \emph{et al.}~\cite{Parasuraman} across four functional types of human-machine interaction, and later adopted in the AI-SEAL framework~\cite{Feldt} to characterize AI support in SE . The LA scale values in Table~\ref{tab:autonomy_levels} describe the LA intervals in which concrete systems operating at a given AL\textsubscript{i} are likely to fall, as an interpretive reference point.

\section{ Illustrative Instantiation of the Metamodel}
\label{sec:instantiation}

\begin{figure*}[htbp]
    \centering
    \includegraphics[width=\textwidth,keepaspectratio]{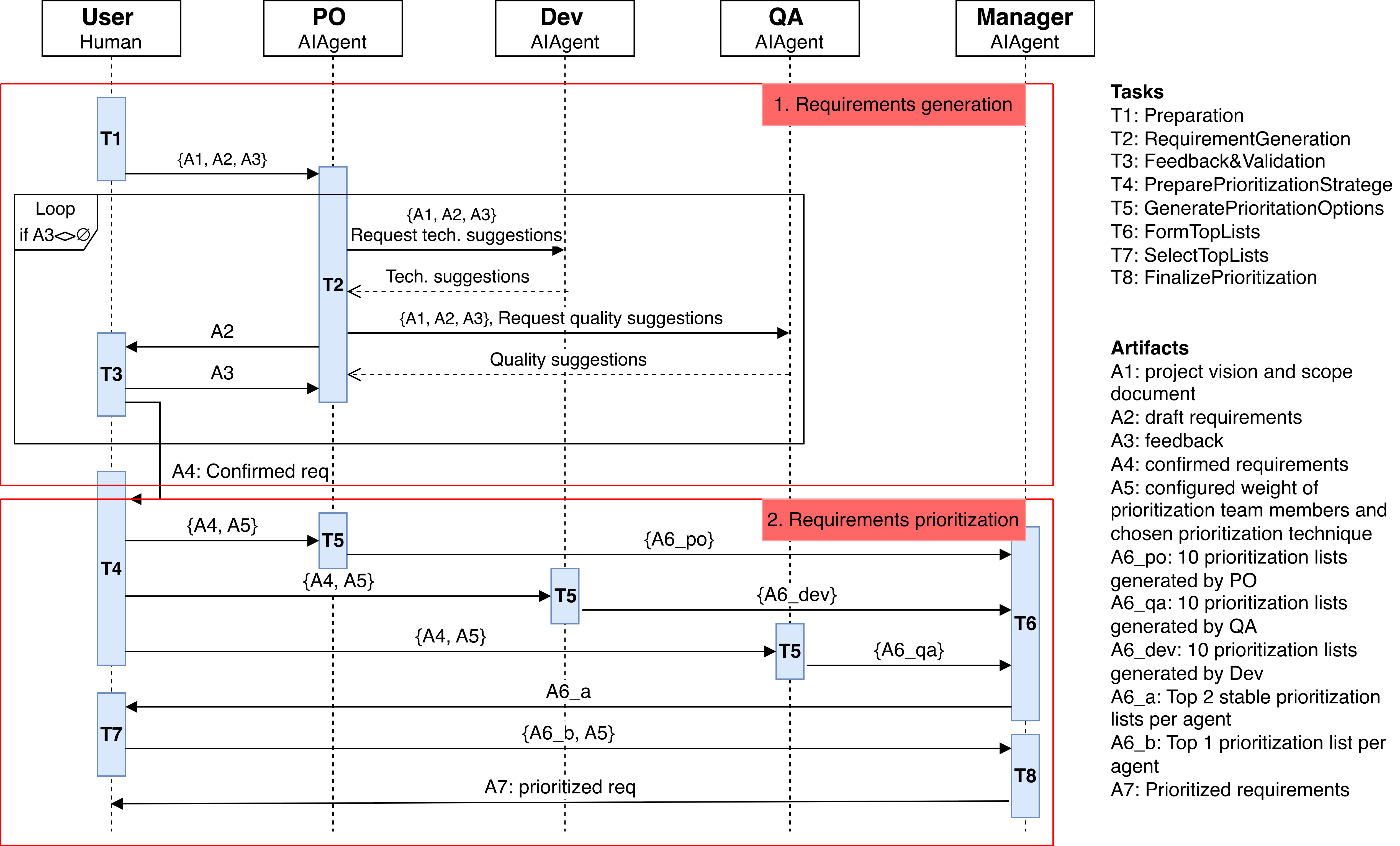}
    \caption{A sequence diagram for the requirements analysis tool}
    \label{fig:req_tool}
\end{figure*}

To illustrate the applicability of the proposed metamodel, we instantiate it using a multi-agent requirements analysis tool \cite{req_tool}. The tool supports user story generation and prioritization through collaboration between a human user and multiple AI agents, including Product Owner (PO), Developer (Dev), Quality Assurance specialist (QA), and Manager agents. The system therefore provides an 
example of a 
human–AI collaboration team conducting requirements analysis tasks
. As shown in Figure ~\ref{fig:req_tool}, the sequence diagram 
captures the interactions among the actors in tasks (T*), as well as the artifacts (A*) they use and produce. The workflow is organized into two collaboration scenarios highlighted in the figure: Requirements Generation and Requirements Prioritization. Both scenarios operate within the Context of requirements engineering and are governed by Constraints such as 
prioritization rules. To keep the presentation clear and simplified, the instantiations of context and constraint, as well as the attributes of all instances are not shown in the diagram, but publicly accessible in the repository\footnote{\url{https://zenodo.org/records/20748480}}.


This tool instantiation illustrates how collaboration can be explicitly modeled through the CollaborationPattern construct. Unlike traditional process models that focus primarily on activities and artifacts, the proposed metamodel captures team structures, decision-making mechanisms, and interaction rules. 
Table~\ref{tab:collaboration_patterns} summarizes two CollaborationPattern instances derived from the tool. Requirements generation is modeled as a hierarchical collaboration pattern with a HumanApproval decision model: the User delegates the task to the PO agent, which coordinates suggestions from Dev and QA before returning the generated requirements for human approval. Requirements prioritization is modeled as a centralized voting pattern: PO, Dev, and QA agents independently evaluate and generate prioritization lists for user stories, while the Manager agent aggregates the results and resolves conflicts according to the User vote.


The instantiation also illustrates the central idea of the metamodel where autonomy is modeled as a dynamic and situation-dependent property. As defined in Section~\ref{sec:AutonomyLevels}, autonomy is not a fixed attribute of an actor. Instead, it is determined as a function of the task being performed, the surrounding context, and the collaboration pattern. The same actor may therefore operate with different autonomy levels across different situations. For example, both collaboration scenarios shown in Figure~\ref{fig:req_tool} 
occur within the same requirements-engineering context. However, differences in autonomy arise from changes in task responsibilities and collaboration patterns. During requirements generation, the PO agent operates at AL2 because it is delegated responsibility for producing the requirements artifact, while the Dev and QA agents operate at AL1 because they only provide supporting suggestions. On the other hand,  during requirements prioritization, the Dev and QA agents independently generate prioritization lists that directly contribute to the final outcome. As a result, their autonomy levels increase from AL1 to AL2. Likewise, human actors may experience varying autonomy depending on organizational policies, regulatory constraints, or safety-critical conditions. 

This analytical instantiation demonstrates that the proposed metamodel not only captures the structural elements of human–AI SE  methods, but also provides explicit mechanisms for modeling collaboration structures, decision-making processes, and dynamic authority redistribution. As a result, the same SE  method can evolve from human-governed collaboration toward increasingly autonomous AI-first teamwork without requiring changes to its underlying process structure.

\begin{table*}[htbp]
\caption{CollaborationPattern instances identified in the requirements analysis tool}
\label{tab:collaboration_patterns}
\centering
\footnotesize
\setlength{\tabcolsep}{4pt}
\renewcommand{\arraystretch}{0.95}
\begin{tabular}{
  p{0.01\textwidth}  
  p{0.10\textwidth}  
  p{0.20\textwidth}  
  p{0.10\textwidth}  
  p{0.50\textwidth}  
}
\toprule
\textbf{id} & \textbf{name} & \textbf{teamStructure} &
\textbf{decisionModel} & \textbf{interactionRules} \\
\midrule
1 &
Requirements Generation &
Hierarchical \newline\{User, PO, Dev, QA\} &
HumanApproval &
User delegates requirements generation to PO; PO requests
suggestions from Dev and QA; Dev and QA return suggestions;
PO synthesizes requirements; User approves or requests revisions. \\
\midrule
2 &
Requirements Prioritization &
Centralized \newline\{User, PO, Dev, QA, Manager\} &
Voting &
User provides the prioritization technique; PO, Dev, and QA
independently prioritize requirements; Manager aggregates
prioritization result based on User vote. \\
\bottomrule
\end{tabular}
\end{table*}

\section{Discussion and Future Plans}
\label{sec:discussion}

The proposed metamodel is intended as an initial step toward autonomy-aware method engineering for human–AI collaboration in AI-first SE  teams. 
Its primary contribution is the separation of method structure from authority assignment. Actors, tasks, artifacts, capabilities, contexts, constraints, and collaboration patterns provide a stable representation of a SE method, while autonomy levels vary according to the task, context, and collaboration pattern. 
The illustrative instantiation further highlights that autonomy should be treated as a situation-dependent property rather than as a fixed characteristic of an actor. Differences in autonomy emerge from task responsibilities, collaboration patterns, governance conditions, and decision-making arrangements, rather than from technical capability alone. This distinction is important for designing AI-first methods that remain adaptable while preserving appropriate human oversight and governance.

This vision paper opens four directions for further research. First, we will \textbf{refine the semantics of the autonomy function} by identifying the concrete task, context, constraint, and collaboration-pattern parameters that influence autonomy assignment. This includes clarifying how team structures, decision models, interaction rules, task criticality, and governance constraints affect the permissible autonomy level of each actor.

Second, we will \textbf{conduct comparative case studies across SE activities} such as requirements analysis, architecture design, coding, testing, and maintenance. These studies will examine whether the same metamodel constructs can represent diverse human–AI collaboration settings and whether additional constructs or relationships are needed.

Third, we will \textbf{develop AI-assisted tool support for automated method modeling}. Such support could extract candidate actors, tasks, artifacts, capabilities, constraints, contexts, and collaboration patterns from process descriptions, workflow traces, or natural-language scenarios, suggest autonomy levels by analyzing task responsibilities, decision models, interaction rules, and governance constraints, and check consistency with governance constraints under human review.

Fourth, we will \textbf{investigate runtime adaptation and governance}. In future AI-first teams, autonomy may need to change dynamically based on task criticality, uncertainty, regulatory conditions, observed performance, or human trust. We therefore plan to explore how the metamodel can be connected to monitoring mechanisms that support adaptive authority allocation, extending it from a design-time specification instrument into a governance mechanism for evolving human–AI SE  teams.

Together, these steps aim to turn the metamodel from a conceptual proposal into an empirically grounded and tool-supported approach for modeling and governing AI-first SE teams.

\section{Conclusion}
\label{sec:conclusion}

This paper proposed an autonomy-aware metamodel for human–AI collaboration in AI-first teams. The metamodel 
treats autonomy as a situation-dependent property derived from the task, context, and collaboration pattern, and formalizes it through authority over task execution, task decomposition, task initiation, and collaboration reconfiguration. By separating method structure from authority assignment, it enables method engineers to model stable collaboration structures while analyzing how authority is distributed across human and AI actors.

The illustrative instantiation with a multi-agent requirements analysis tool shows how autonomy levels may vary across requirements generation and prioritization according to task responsibilities, team structure, decision model, and interaction rules. As a vision paper, this work points toward a new generation of software engineering methods that are not only AI-enabled, but also governable, adaptable, and explicit about authority distribution in human–AI collaboration.




\begin{acks}
This research was partially supported by the ANSE (AI Native Software Engineering) project funded by Business Finland and by FAST (Finnish Software Engineering Doctoral Research Network). The authors gratefully acknowledge this support.
\end{acks}

\printbibliography

@String{Computing = "Computing" }

@String{Computer = "{IEEE} Computer" }

@String{Springer = "Springer-Verlag" }

@article{fuggetta2002classification,
  title={A classification of CASE technology},
  author={Fuggetta, Alfonso},
  journal={Computer},
  volume={26},
  number={12},
  pages={25--38},
  year={2002},
  publisher={IEEE}
}

@article{brinkkemper1996method,
  title={Method engineering: engineering of information systems development methods and tools},
  author={Brinkkemper, Sjaak},
  journal={Information and software technology},
  volume={38},
  number={4},
  pages={275--280},
  year={1996},
  publisher={Elsevier}
}

@article{hassan2024towards,
  title={Towards AI-native software engineering (SE 3.0): A vision and a challenge roadmap},
  author={Hassan, Ahmed E and Oliva, Gustavo A and Lin, Dayi and Chen, Boyuan and Ming, Zhen and others},
  journal={arXiv preprint arXiv:2410.06107},
  year={2024}
}

@article{lu2021design,
  title={Design ontology supporting model-based systems engineering formalisms},
  author={Lu, Jinzhi and Ma, Junda and Zheng, Xiaochen and Wang, Guoxin and Li, Han and Kiritsis, Dimitris},
  journal={IEEE Systems Journal},
  volume={16},
  number={4},
  pages={5465--5476},
  year={2021},
  publisher={IEEE}
}

@inproceedings{hong2023metagpt,
  title={MetaGPT: Meta programming for a multi-agent collaborative framework},
  author={Hong, Sirui and Zhuge, Mingchen and Chen, Jonathan and Zheng, Xiawu and Cheng, Yuheng and Wang, Jinlin and Zhang, Ceyao and Wang, Zili and Yau, Steven Ka Shing and Lin, Zijuan and others},
  booktitle={The twelfth international conference on learning representations},
  year={2023}
}

@article{ge2025survey,
  title={A survey of vibe coding with large language models},
  author={Ge, Yuyao and Mei, Lingrui and Duan, Zenghao and Li, Tianhao and Zheng, Yujia and Wang, Yiwei and Wang, Lexin and Yao, Jiayu and Liu, Tianyu and Cai, Yujun and others},
  journal={arXiv preprint arXiv:2510.12399},
  year={2025}
}

@article{OMG2008,
    title={Software \& Systems Process Engineering Metamodel Specification (SPEM) Version 2.0},
  author={OMG},
  journal={Tech. rep.},
  year={2008},
  publisher={Object Management Group}
}

@article{sommerville2016software,
  title={Software Engineering 10th Edition (International Computer Science)},
  author={Sommerville, I},
  journal={Essex, UK: Pearson Education},
  pages={1--808},
  year={2016}
}

@article{henderson2010situational,
  title={Situational method engineering: state-of-the-art review},
  author={Henderson-Sellers, Brian and Ralyt{\'e}, Jolita},
  journal={Journal of Universal Computer Science},
  year={2010}
}

@article{bella2023ontology,
  title={The ontology for agents, systems and integration of services: OASIS version 2},
  author={Bella, Giampaolo and Cantone, Domenico and Longo, Carmelo Fabio and Nicolosi-Asmundo, Marianna and Santamaria, Daniele Francesco},
  journal={Intelligenza Artificiale},
  volume={17},
  number={1},
  pages={51--62},
  year={2023},
  publisher={SAGE Publications Sage UK: London, England}
}

@article{jacobson2012essence,
  title={The essence of software engineering: the SEMAT kernel},
  author={Jacobson, Ivar and Ng, Pan-Wei and McMahon, Paul E and Spence, Ian and Lidman, Svante},
  journal={Communications of the ACM},
  volume={55},
  number={12},
  pages={42--49},
  year={2012},
  publisher={ACM New York, NY, USA}
}

@inproceedings{gonzalez2007supporting,
  title={Supporting situational method engineering with ISO/IEC 24744 and the work product pool approach},
  author={Gonzalez-Perez, Cesar},
  booktitle={Working conference on method engineering},
  pages={7--18},
  year={2007},
  organization={Springer}
}

@InProceedings{req_tool,
author="Sami, Malik Abdul
and Zhang, Zheying
and Waseem, Muhammad
and Kemell, Kai-Kristian
and Rasheed, Zeeshan
and Herda, Tomas
and Hasan, Md. Toufique
and Rasku, Jussi
and Abrahamsson, Pekka",
editor="Taibi, Davide
and Smite, Darja",
title="A Multi-agent LLM System for Automated Requirements Analysis: A Study on User Story Generation and Prioritization",
booktitle="Software Engineering and Advanced Applications",
year="2026",
publisher="Springer Nature Switzerland",
address="Cham",
pages="178--187",
isbn="978-3-032-04200-2"
}

@article{RobotAutonomy,
author = {Beer, Jenay M. and Fisk, Arthur D. and Rogers, Wendy A.},
title = {Toward a framework for levels of robot autonomy in human-robot interaction},
year = {2014},
issue_date = {July 2014},
publisher = {Journal of Human-Robot Interaction Steering Committee},
volume = {3},
number = {2},
url = {https://doi.org/10.5898/JHRI.3.2.Beer},
doi = {10.5898/JHRI.3.2.Beer},
journal = {J. Hum.-Robot Interact.},
month = jul,
pages = {74–99},
numpages = {26}
}

@article{HumanAutonomy,
  author  = {O'Neill, Thomas A. and McNeese, Nathan J. and Barron, Ashley and Schelble, Brandon},
  title   = {Human--Autonomy Teaming: A Review and Analysis of the Empirical Literature},
  journal = {Human Factors},
  year    = {2022},
  volume  = {64},
  number  = {5},
  pages   = {904--938},
  doi     = {10.1177/0018720820960865},
  url     = {https://doi.org/10.1177/0018720820960865}
}

@inproceedings{inproceedingsHarmsen1994,
author = {Harmsen, Frank and Brinkkemper, Sjaak and Oei, J.},
year = {1994},
month = {01},
pages = {169-194},
title = {Situational method engineering for information system project approaches},
journal = {IFIP Transactions A: Computer Science and Technology}
}

@article{musick2021happens,
  title={What happens when humans believe their teammate is an AI? An investigation into humans teaming with autonomy},
  author={Musick, Geoff and O'Neill, Thomas A and Schelble, Beau G and McNeese, Nathan J and Henke, Jonn B},
  journal={Computers in Human Behavior},
  volume={122},
  pages={106852},
  year={2021},
  publisher={Elsevier}
}

@article{karakama2021task,
  title={Task Decomposition and Role Sharing for Real-time Human-AI Swarm Collaboration},
  author={Karakama, Sotaro and Matsunami, Natsuki and Ito, Masayuki},
  journal={International Journal of Smart Computing and Artificial Intelligence},
  volume={5},
  number={1},
  pages={33--50},
  year={2021}
}

@article{naik2025exploring,
  title={Exploring Human-AI Collaboration Using Mental Models of Early Adopters of Multi-Agent Generative AI Tools},
  author={Naik, Suchismita and Toombs, Austin L and Snellinger, Amanda and Saponas, Scott and Hall, Amanda K},
  journal={arXiv preprint arXiv:2510.06224},
  year={2025}
}

@article{bauer2004methodologies,
  title={Methodologies and modeling languages},
  author={Bauer, Bernhard and M{\"u}ller, J{\"o}rg P},
  journal={Agent-Based Software Development},
  pages={77--131},
  year={2004},
  publisher={Artech House London}
}

@article{qiang2023agent,
  title={Agent-om: Leveraging llm agents for ontology matching},
  author={Qiang, Zhangcheng and Wang, Weiqing and Taylor, Kerry},
  journal={arXiv preprint arXiv:2312.00326},
  year={2023}
}

@article{Sheridan,
author = {Sheridan, Thomas and Verplank, W. and Brooks, T.},
year = {1978},
month = {12},
pages = {},
title = {Human and Computer Control of Undersea Teleoperators}
}

@ARTICLE{Parasuraman,
  author={Parasuraman, R. and Sheridan, T.B. and Wickens, C.D.},
  journal={IEEE Transactions on Systems, Man, and Cybernetics - Part A: Systems and Humans}, 
  title={A model for types and levels of human interaction with automation}, 
  year={2000},
  volume={30},
  number={3},
  pages={286-297},
  doi={10.1109/3468.844354}}

@inproceedings{Feldt,
author = {Feldt, Robert and de Oliveira Neto, Francisco G. and Torkar, Richard},
title = {Ways of applying artificial intelligence in software engineering},
year = {2018},
isbn = {9781450357234},
publisher = {Association for Computing Machinery},
address = {New York, NY, USA},
url = {https://doi.org/10.1145/3194104.3194109},
doi = {10.1145/3194104.3194109},
booktitle = {Proceedings of the 6th International Workshop on Realizing Artificial Intelligence Synergies in Software Engineering},
pages = {35–41},
numpages = {7},
location = {Gothenburg, Sweden},
series = {RAISE '18}
}

@INPROCEEDINGS{ait2025towards,
  author={Ait, Adem and Jouneaux, Gwendal and Cánovas Izquierdo, Javier Luis and Cabot, Jordi},
  booktitle={2025 40th IEEE/ACM International Conference on Automated Software Engineering (ASE)}, 
  title={Towards Automated Governance: A DSL for Human-Agent Collaboration in Software Projects}, 
  year={2025},
  volume={},
  number={},
  pages={3891-3895},
  doi={10.1109/ASE63991.2025.00337}}

@InProceedings{ait2026,
author="Ait, Adem
and Izquierdo, Javier Luis C{\'a}novas
and Cabot, Jordi",
editor="Taibi, Davide
and Smite, Darja",
title="Towards Modeling Human-Agentic Collaborative Workflows: A BPMN Extension",
booktitle="Software Engineering and Advanced Applications",
year="2026",
publisher="Springer Nature Switzerland",
address="Cham",
pages="367--382",
}

@article{liu2025agent,
  title={Agent design pattern catalogue: A collection of architectural patterns for foundation model based agents},
  author={Liu, Yue and Lo, Sin Kit and Lu, Qinghua and Zhu, Liming and Zhao, Dehai and Xu, Xiwei and Harrer, Stefan and Whittle, Jon},
  journal={Journal of Systems and Software},
  volume={220},
  pages={112278},
  year={2025},
  publisher={Elsevier}
}

@article{MGS-2012,
author = {K\"{u}ster, Tobias and L\"{u}tzenberger, Marco and He\ss{}ler, Axel and Hirsch, Benjamin},
title = {Integrating process modelling into multi-agent system engineering},
year = {2012},
issue_date = {January 2012},
publisher = {IOS Press},
address = {NLD},
volume = {8},
number = {1},
issn = {1574-1702},
url = {https://doi.org/10.3233/MGS-2012-0182},
doi = {10.3233/MGS-2012-0182},
month = jan,
pages = {105–124},
numpages = {20}
}

@InProceedings{10.1007/978-3-032-12089-2_26,
author="Siddeeq, Shahbaz
and Waseem, Muhammad
and Rasheed, Zeeshan
and Hasan, Md Mahade
and Rasku, Jussi
and Saari, Mika
and Terho, Henri
and M{\"a}kel{\"a}, Kalle
and Kemell, Kai-Kristian
and Abrahamsson, Pekka",
title="LLM-Based Multi-agent System for Intelligent Refactoring of Haskell Code",
booktitle="Product-Focused Software Process Improvement",
year="2026",
publisher="Springer Nature Switzerland",
pages="408--418",
}
\end{document}